\documentclass[12pt]{article}

\usepackage{makecell}
\usepackage{amsmath}
\usepackage{amssymb}
\usepackage{slashed}
\usepackage{amsfonts}
\usepackage{graphicx}
\usepackage{soul}
\usepackage{pdflscape}
\usepackage[dvipsnames]{xcolor}
\usepackage[colorlinks=true,urlcolor=blue,linkcolor=blue,citecolor=blue]{hyperref}
\usepackage{cancel}
\usepackage{soul}
\usepackage{booktabs}
\usepackage{caption}
\usepackage{subcaption}
\usepackage{float}

\usepackage[top=1.135in,bottom=1.17in,left=1.16in,right=1.16in]{geometry}
\numberwithin{equation}{section}
\allowdisplaybreaks

\newcommand{\be}{\begin{equation}}
\newcommand{\ee}{\end{equation}}

\newcommand{\eq}[1]{\begin{align}#1\end{align}}

\usepackage{graphicx} 
\usepackage{amssymb,amsthm,amsmath,amsfonts}

\usepackage{makecell}
\usepackage{bm}
\usepackage{multicol}
\usepackage{appendix}
\usepackage{orcidlink}

\usepackage[T1]{fontenc}

\begin{document}
\thispagestyle{empty}

\begin{center}

{\bf\Large \boldmath Proton-box contribution to $a_{\mu}^{\rm{HLbL}}$}

\vspace{50pt}

Emilio J. Estrada~\orcidlink{0009-0007-3360-7908}$^{1}$, Juan Manuel Márquez~\orcidlink{0009-0007-3354-2497} $^{1}$, Diego Portillo-Sánchez~\orcidlink{0000-0002-3354-6355}$^{1}$ and Pablo Roig~\orcidlink{0000-0002-6612-7157} $^{1,2}$

\vspace{16pt}
{$^{1}$\it Departamento de F\'isica, Centro de Investigaci\'on y de Estudios Avanzados del Instituto Polit\'ecnico Nacional} \\
{\it Apartado Postal 14-740, 07360 Ciudad de México, M\'exico}\\
{$^{2}$\it IFIC, Universitat de Val\`encia – CSIC, Catedr\'atico Jos\'e Beltr\'an 2, E-46980 Paterna, Spain} \\
\vspace{16pt}

\vspace{50pt}

\vspace{16pt}
{\small
}
\vspace{16pt}

{\tt}

\vspace{30pt}

\end{center}

\begin{abstract}
We analyze the proton$\text{-}$box contribution to the hadronic light$\text{-}$by$\text{-}$light part of the muon's anomalous magnetic moment, which is the first reported baryonic contribution to this piece.
We follow the quark$\text{-}$loop analysis, incorporating the relevant data$\text{-}$driven and lattice proton form factors. Although the heavy mass expansion would yield a contribution of $\mathcal{O}(10^{-10})$, the damping of the form factors in the regions where the kernel peaks, explains our finding $a_{\mu}^{\rm{p-box}}=1.82 (7)\times 10^{-12}$, two orders of magnitude smaller than the forthcoming uncertainty on the $a_{\mu}$ measurement and on its Standard Model prediction.
\end{abstract}

\newpage


\section{Introduction}
The anomalous magnetic moment of the muon has attracted significant attention in particle physics due to the persistent discrepancy between its experimental measurement and the theoretical predictions of the Standard Model (SM). The latest measurements by the Muon $g$-2 collaboration at Fermilab \cite{Muong-2:2023cdq,Muong-2:2021ojo}, when combined with earlier results from the Brookhaven E821 experiment \cite{Muong-2:2006rrc}, indicate a deviation of 5.1$\sigma$ from the SM prediction \cite{Aoyama:2020ynm} (which is based on refs.~\cite{Davier:2017zfy,Keshavarzi:2018mgv,Colangelo:2018mtw,Hoferichter:2019mqg,Davier:2019can,Keshavarzi:2019abf,Kurz:2014wya,FermilabLattice:2017wgj,Budapest-Marseille-Wuppertal:2017okr,RBC:2018dos,Giusti:2019xct,Shintani:2019wai,FermilabLattice:2019ugu,Gerardin:2019rua,Aubin:2019usy,Giusti:2019hkz,Melnikov:2003xd,Masjuan:2017tvw,Colangelo:2017fiz,Hoferichter:2018kwz,Gerardin:2019vio,Bijnens:2019ghy,Colangelo:2019uex,Pauk:2014rta,Danilkin:2016hnh,Jegerlehner:2017gek,Knecht:2018sci,Eichmann:2019bqf,Roig:2019reh,Colangelo:2014qya,Blum:2019ugy,Aoyama:2012wk,Aoyama:2019ryr,Czarnecki:2002nt,Gnendiger:2013pva}):
\begin{equation} \label{eq:delta_a_mu}
\Delta a_\mu= a_\mu^{\text{exp}}-a_\mu^{\text{SM}}=
249(48)\times 10^{-11}.
\end{equation}
Furthermore, there is a discrepancy among theoretical predictions, specifically in the Hadronic Vacuum Polarization (HVP) determination, which can be derived using different approaches. The first method utilizes $e^{+}e^{-}$ data-driven techniques, yielding the previously mentioned 5.1$\sigma$ tension. In contrast, estimates based on $\tau$ data-driven approaches \cite{Miranda:2020wdg,Masjuan:2024mlg,Davier:2023fpl} or lattice QCD calculations \cite{Borsanyi:2020mff} significantly reduce the tension between theoretical and experimental values to 2.0$\sigma$ and 1.5$\sigma$, respectively (less than one $\sigma$ in \cite{Boccaletti:2024guq}). The latest CMD-3 measurement of $\sigma(e^+e^-\to\pi^+\pi^-)$ \cite{CMD-3:2023alj,CMD-3:2023rfe} also points in this direction.
 
As the uncertainty in $a_\mu^{\text{exp}}$ is expected to decrease further as more data is analyzed,~\footnote{With the FNAL Run-2/3 analysis\cite{Muong-2:2023cdq} $a_\mu^{\mathrm{exp}}$ was measured with a precision of 0.20 ppm, and a 0.14 ppm is expected to be achieved when the already taken data of Run-4/5/6 analysis is finished. Besides, the J-PARC E34\cite{Lee:2019xkg} aims to measure it using a different approach than BNL E821\cite{Muong-2:2006rrc} and FNAL, with totally different systematic uncertainties.} a significant improvement of the theoretical calculation is essential to determine whether this discrepancy can be attributed to New Physics (NP). Therefore, its different contributions are continuously refined to achieve this objective, particularly the hadronic ones (HVP and hadronic light-by-light (HLbL)), which dominate the uncertainty.~\footnote{The precision of QED and Electroweak determinations are two and one order of magnitude more accurate than the hadronic ones, respectively.} The HVP data-driven computation is directly related to the experimental input from $\sigma(e^+e^-\to \mathrm{hadrons})$ data. HLbL in contrast, requires a decomposition in all possible intermediate states.

Recently, a rigorous framework, based on the fundamental principles of unitarity, analyticity, crossing symmetry,
and gauge invariance has been developed  \cite{Colangelo:2015ama,Colangelo:2017fiz}, providing a clear and precise methodology for defining and evaluating  the various low-energy contributions to HLbL scattering. The most significant among these are the pseudoscalar-pole ($\pi^0$, $\eta$ and $\eta^\prime$) contributions\cite{Masjuan:2017tvw,Hoferichter:2018kwz,Hoferichter:2018dmo,Estrada:2024cfy}. Nevertheless, subleading pieces, such as the $\pi^{\pm}$ and $K^{\pm}$ box diagrams, along with quark loops, have also been reported \cite{Colangelo:2019uex,Eichmann:2019bqf,Miramontes:2021exi,Stamen:2022uqh}, with the proton-box representing an intriguing follow-up calculation.  We emphasize that the computation of $a_{\mu}^{\rm{HLbL}}$ in the present work belongs to the dispersive evaluation, which allows us to use the same theoretical framework developed in Refs.\cite{Colangelo:2014qya,Colangelo:2015ama,Colangelo:2019uex}, with a specific focus in the proton-box diagrams, that adds up to existing results in the dispersive framework. Other analogous baryon-box contributions to $a_\mu^{\text HLbL}$ are also of interest, nonetheless. However, as we shall elaborate later, these contributions are expected to be subdominant compared with the one coming from the proton-box. Then, we concentrate our efforts on estimating the first baryon box-loop contribution to $a_{\mu}^{\rm{HLbL}}$, leaving its generalization for the rest of the baryon-octet to future work.

Specifically, a preliminary result obtained from the Heavy Mass Expansion (HME) method \cite{Kuhn:2003pu}—which does not consider the form factors contributions— for a mass of $M\equiv M_p=938$ MeV, yields an approximate mean value of $a_{\mu}^{\mathrm{p-box}}=9.7\times10^{-11}$ \footnote{{ In the same fashion, we can also estimate $a_{\mu}^{\rm{\Sigma}-box}\approx 6\times 10^{-11}$, $a_{\mu}^{\rm{\Xi-box}}\approx 4.9\times 10^{-11}$ using the HME method as a rough estimation, which are smaller than the proton contribution, as expected.}}. This result is comparable in magnitude to several of the previously discussed contributions, thereby motivating a more realistic and precise analysis that incorporates the main effects of the relevant form factors.

In this work, we focus on the proton-box HLbL contribution. We apply the master formula and the perturbative quark loop scalar functions, derived in \cite{Colangelo:2015ama} (which we verified independently), together with a complete analysis of different proton form factors descriptions\cite{Alexandrou:2018sjm,Alberico:2008sz,Ye:2017gyb}, which are essential inputs for the numerical integration required in the calculations.

This paper is structured as follows: after a review of the master formula needed to compute the possible HLbL contribution to $a_\mu$ in section \ref{sec:2}, we discuss the proton form factors, required for the analysis, in full detail in section \ref{sec:3}. Then, in section \ref{sec:4}, we report the first evaluation of the proton-box HLbL contribution and the associated error estimate. Finally, our conclusions are given in section \ref{sec:5}.

\section{HLbL Master Formula}
\label{sec:2}

\begin{figure}[htp]
    \centering
    \includegraphics[width=0.35\linewidth]{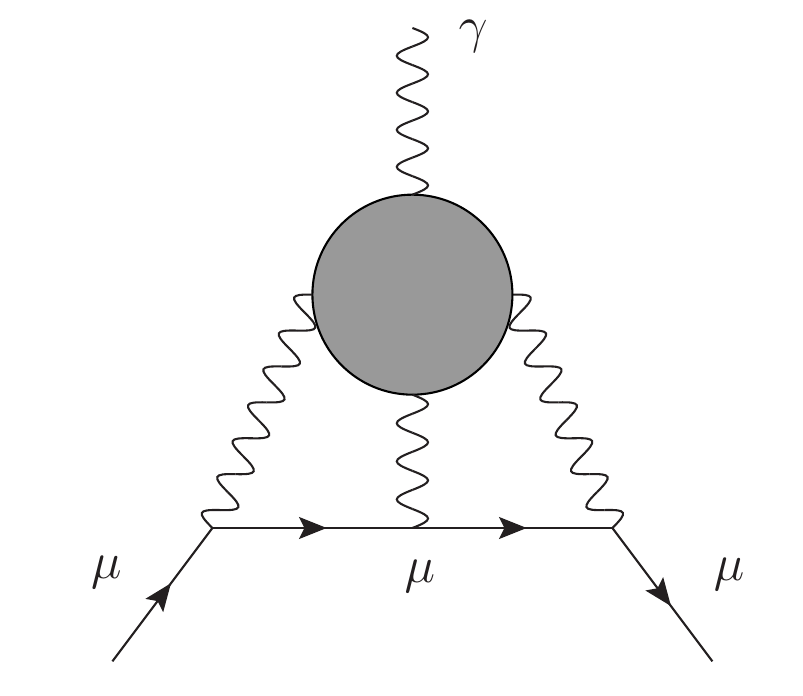}
    \caption{General light-by-light contribution to the muon g-2 anomaly.}
    \label{fig:light-by-light}
\end{figure}

A comprehensive review of the HLbL contributions to the muon $g$-2 anomaly has been previously done in Refs.\cite{Colangelo:2015ama,Colangelo:2017fiz, Colangelo:2019uex}, where the authors obtained a master integral,  allowing to use the unitarity relation to consider individual intermediate states and account for their contribution to $a_{\mu}^{\rm{HLbL}}$ using physical observables, such as on-shell form factors -which can be given by a model, lattice calculations or parametrizations of experimental data-, for a general description of the electromagnetic tensor involved in the the two-loop diagram shown in Fig.\ref{fig:light-by-light}, leading to:

\eq{
a_{\mu}^{\rm{HLbL}}=\frac{2\alpha^3}{3\pi^2}\int_0^{\infty}dQ_1\int_0^{\infty}dQ_2\int_{-1}^{1}d\tau \sqrt{1-\tau^2}\,Q_1^3\,Q_2^3\sum_{i=1}^{12}T_i(Q_1,Q_2,\tau)\bar{\Pi}_i(Q_1,Q_2,\tau), \label{eq:master-formula}}

where $Q_i^2=-q_i^2$ is the square four-momentum of the photons in the space-like region and $\tau$ is defined by the relation $Q_3^2=Q_1^2+Q_2^2+2 Q_1 Q_2 \tau$. Moreover, $T_i$ are the 12 independent integral kernels (explicitly shown in Appendix B of Ref.\cite{Colangelo:2017fiz}) and $\bar{\Pi}_i$ corresponds to scalar functions that encode all the information of the specific HLbL intermediate state contribution. Both can be obtained after a decomposition of the light-by-light tensor, following the recipe introduced by Bardeen-Tung-Tarrach (BTT) \cite{Bardeen:1968ebo,Tarrach:1975tu}.~\footnote{This procedure ensures that the expressions are free of kinematic zeros and singularities, see \cite{Colangelo:2015ama,Colangelo:2017fiz} for details.} Indeed, analytical expressions for $\bar{\Pi}_i$ have been computed for different contributions, such as the pseudo-scalar poles, pseudo-scalar and fermion box-diagrams, etc. \cite{Colangelo:2015ama,Colangelo:2017fiz,Colangelo:2019uex}.

\begin{figure}[htp]
    \centering
    \begin{tabular}{ccc}
        \includegraphics[scale=.33]{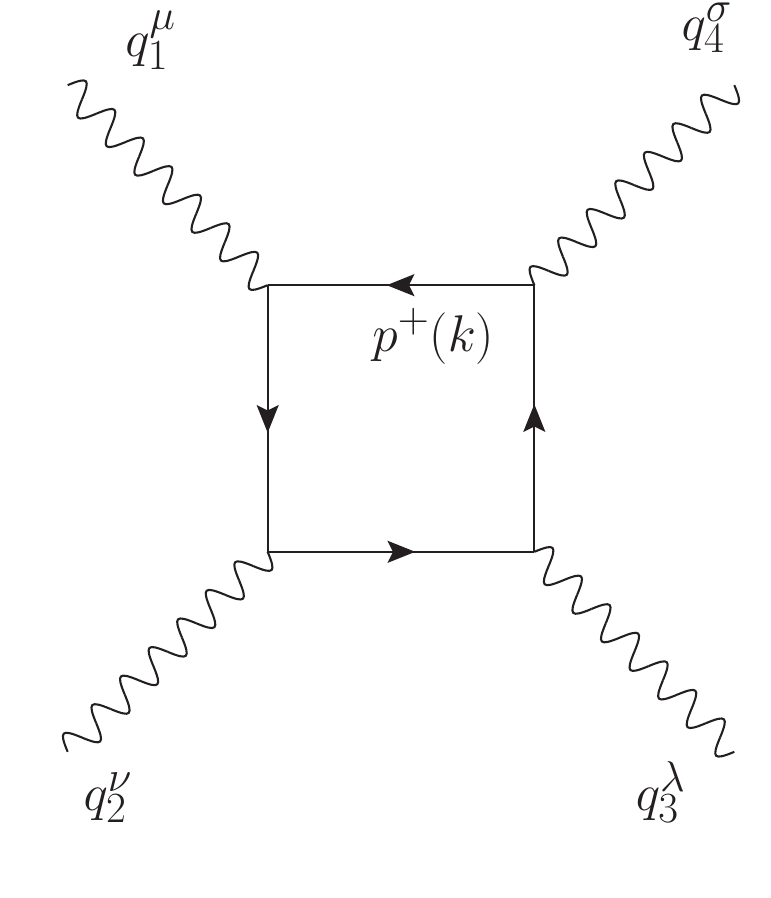} & \includegraphics[scale=.33]{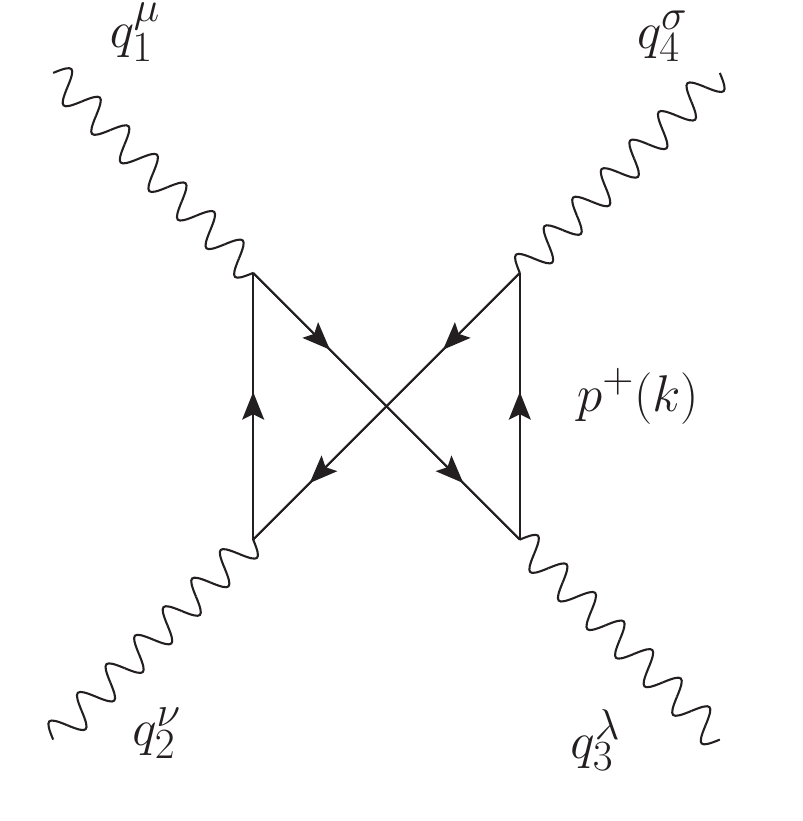} & \includegraphics[scale=.33]{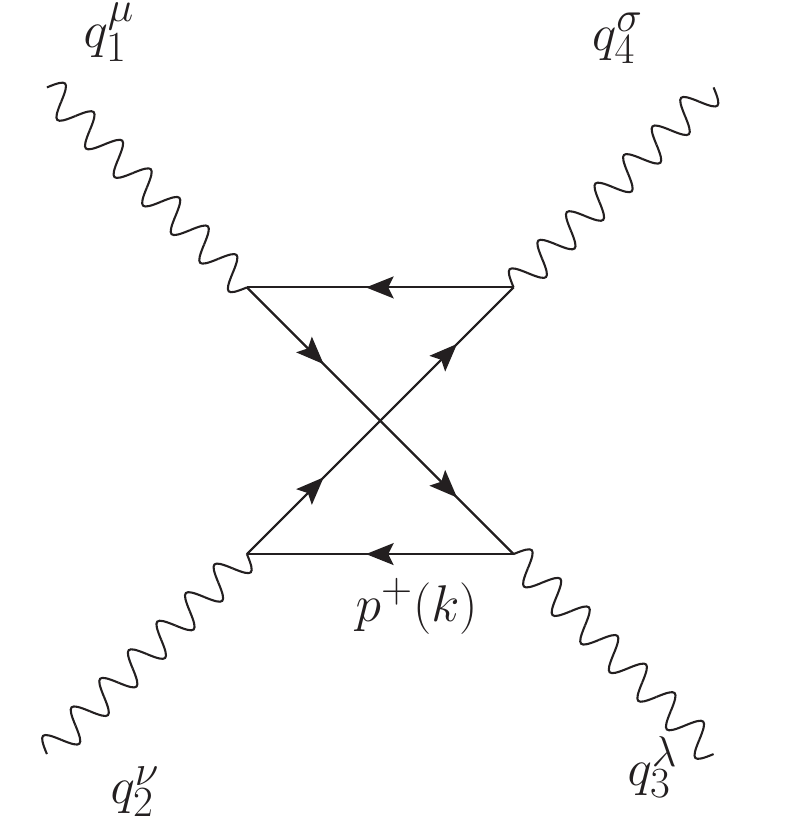} \\
    \end{tabular}
    \caption{Feynman diagrams for light-by-light scattering induced by a proton-box loop (and the corresponding diagrams with exchanged fermion fluxes inside the loop).}
    \label{fig:diagrams}
\end{figure}

The aim of this work is to first evaluate the proton-box contribution (Fig.\ref{fig:diagrams}) to $a_{\mu}^{\rm{HLbL}}$. For this purpose, we consider the following general nucleon-photon matrix element, consistent with Lorentz and gauge invariance, as well as $\mathcal{P}$ and $\mathcal{C}\mathcal{P}$ conservation:
\eq{
\langle p^+(P_2)|J^{\mu}_{e.m.}(q)|p^{+}(P_1)\rangle&=\bar{u}(P_2)\,\Gamma^{\mu}(q)u(P_1)\\
&=\bar{u}(P_2)\left(F_1(q^2) \gamma^{\mu}+i\,\frac{F_2(q^2)}{2\,M_p}\sigma^{\mu\,\nu}q_{\nu}\right)u(P_1),\label{eq:matrix-element}}
where $F_{1,2}$ are the proton Dirac and Pauli form factors, respectively, and $q=P_2-P_1$.

In this analysis, the dominant term in eq.(\ref{eq:matrix-element}), is the one proportional to $F_1(k^2)$. This arises from the additional $q_{\nu}/M_p$ factor appearing in the tensor vertex, in such a way that, in the low momentum transfer regime (below $1$ GeV), the tensor coupling behavior is significantly suppressed due to its momentum dependence, making it a valid approximation to consider only the vector coupling.\footnote{This simplification is commonly taken, consistently in many other processes, such as electron-proton scattering in the same momentum transfer region. Similarly, in the proton-loop contribution to HVP, incorporating a non-zero $F_2$ leads to a 0.02$\%$ modification of the central value, which remains significantly smaller than the current uncertainty (consistent with the results reported in ref. \cite{Keshavarzi:2019abf}).} Conversely, at high photon virtuality (or even above $1$ GeV), the $F_2(q^2)$ behavior becomes the suppression factor. In fact, in this case, also the  $F_1(q^2)$ and kernel functions are highly suppressed, making the contributions of this $q^2$ region negligible.  Indeed, due to the asymptotic constrains of the form factors $F_1$ and $F_2$ from p-QCD\cite{Brodsky:1973kr}, which should behave as $\sim Q^{-4},Q^{-6}$, respectively (satisfied by construction in both parametrizations employed in this work, as we discuss latter), the regime of high transferred momentum is free of divergences and we do not expect to have any significant error coming from this approximation \footnote{{In this sense, we have: $\Gamma^{\mu}(q)|{}_{q\rightarrow\infty}= \gamma^{\mu}\, A\,\,q^{-4} + i \,B\,\sigma^{\mu\nu} \hat{q}_{\nu}(2\, q^5 \,M_p )^{-1}$, with $\hat{q}_{\nu}\equiv q_{\nu}/q$ a normalized four-vector and $A$ and $B$ being constants.}}. 

Therefore, as a suitable first approximation, we will work only with the vector coupling in eq.(\ref{eq:matrix-element}) as input for the scattering amplitude computation shown in Fig. \ref{fig:diagrams}. These considerations lead to the following form of the scalar functions required in the $a_{\mu}^{\rm{HLbL}}$ master integral:
\eq{
\bar{\Pi}_i=F_1(Q_1^2)F_1(Q_2^2)F_1(Q_3^2)\,\frac{1}{16\pi^2}\int_0^1 dx \int_0^{1-x} dy\, I_i(Q_1,Q_2,\tau,x,y),\label{eq:Scalar-functions}
}
where, for completeness, we show the $I_i$ functions in appendix \ref{App:Scalar-functions}. The only difference between the above expression and the one reported in \cite{Colangelo:2019uex} (for the quark-box loop) are the presence of the vector form factors $F_1(Q^2)$ and the absence of the $N_C Q^4_q$ global quark-factor in the Feynman parameter integrals in eq.~(\ref{eq:Scalar-functions}). 

\begin{figure}[htp]
\centering
\includegraphics[scale=.23]{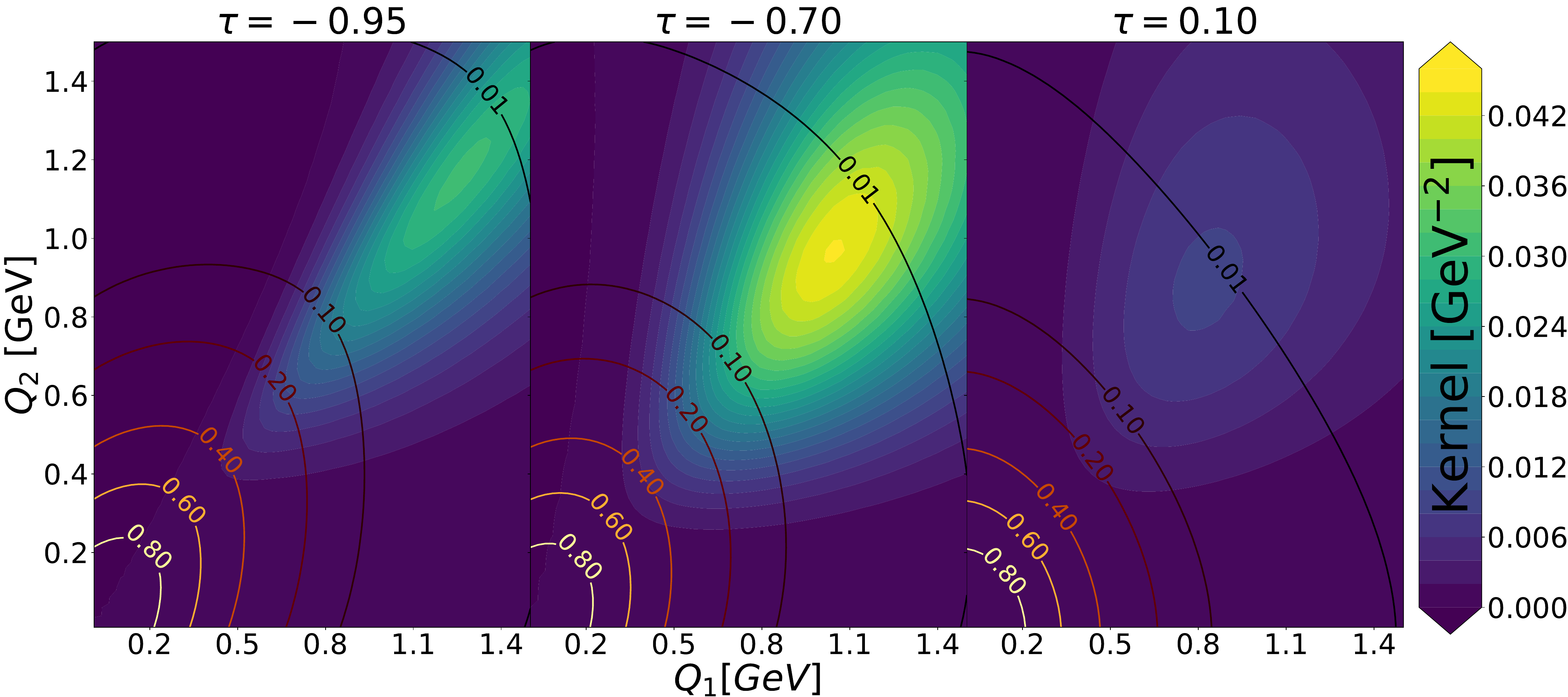}\caption{Integral kernel (density plot) versus form factors dependence (contour plot) at relevant different virtualities for the off-shell photons, as described in the text.}
    \label{fig:kernels}
\end{figure}

A preliminary analysis reveals that the full integral kernel $\sqrt{1-\tau^2}\,Q_1^3\,Q_2^3 T_i\bar{\Pi}_i$,~\footnote{The explicit dependence of $T_i$ and $\bar{\Pi}_i$ on $Q_1,Q_2$ and $\tau$ has been omitted and the Einstein sum notation is understood.} without accounting for any form factor, primarily contributes to the overall integral at low momentum transfers (below $1$ GeV), as illustrated by the density plot in Fig.~\ref{fig:kernels} for three different $\tau$ values. Furthermore, the vector proton form factors term alone $F_1(Q_1^2)F_1(Q_2^2)F_1(Q_3^2)$,~\footnote{Evaluated using the dependence on $Q^2$ as a z-expansion (setup 1), see section \ref{sec:3} for details.} acquires its maximum value within the same momenta region, as shown by the contour curves in the figure. 
Finally, due to the mismatch of the kernels and form factors maximum values, a significant decrease of the HME approximation result is expected to happen. We will delve deeper into this in following sections.

\section{Proton Form Factors}
\label{sec:3}

In order to perform an accurate estimation of the first main contribution of the proton-box to $a_{\mu}^{\rm{HLbL}}$, we compute numerically the master integral in eq.(\ref{eq:master-formula}) taking into account two different descriptions of the form factors. The first one will be a data-driven approach~\cite{Ye:2017gyb}, followed by a lattice QCD computation~\cite{Alexandrou:2018sjm}. Both approaches fitted parametrizations to the electric ($G_E$) and magnetic ($G_M$) proton form factors data, which are related to the Dirac and Pauli basis by:
\eq{
G_{E}(Q^2)&=F_1(Q^2)-\frac{Q^2}{4 M_p^2}\,F_2(Q^2),\\
G_{M}(Q^2)&=F_1(Q^2)+F_2(Q^2).
} 
\subsection{Setup 1: Data-Driven Form Factors}
First, we make use of the electric and magnetic form factors obtained in Ref.\cite{Ye:2017gyb} after fitting the experimental data to a z-expansion parametrization of order 12 \cite{Hill:2010yb}, where sum-rule constraints were applied on each form factor to warrant the asymptotic scaling $G_{E,M} \sim Q^{-4}$ and the correct normalization at null photon virtuality.~\footnote{Other parametrizations, as the ones reported in Refs.\cite{Alberico:2008sz,Arrington:2007ux}, have been considered for the $a_\mu^\mathrm{p-box}$ numerical evaluation, being consistent with the one used in this work within less than 1$\sigma$.} Both systematic and statistical errors were addressed in the computation of this form factor, which was implemented in our work. In this way, the proton form factors can be written as:
\eq{
G_{E}^{(p)}(Q^2),\,\frac{G_{M}^{(p)}(Q^2)}{\mu_p}=\sum_{i=0}^{12}a_i^{\{E,M\}}z^i,\label{eq:formfactors}
}
where $a_i$ are fitting parameters shown in table \ref{tab:parameters}, and $z$ is defined as follows: 
\eq{
z\equiv\frac{\sqrt{t_{\rm{cut}}+Q^2}-\sqrt{t_{\rm{cut}}-t_0}}{\sqrt{t_{\rm{cut}}+Q^2}+\sqrt{t_{\rm{cut}}-t_0}},
} with $t_0=-0.7$ GeV${}^2$, $t_{\rm{cut}}=4 m_{\pi}^2$ and
the form factors normalization fixed by the proton's electric charge non-renormalization and magnetic moment in Bohr magneton units, $G_E^p(0)=1$ and $G_M^p(0)=\mu_p=2.793$, in turn. 
\begin{table}[htp]
    \centering
    \begin{tabular}{c|c|c}\hline\hline
              & E & M \\\hline\hline 
        $a_0^X$ & $0.239163298067$ & $0.264142994136$ \\
        $a_1^X$ & $-1.109858574410$ & $-1.095306122120$\\
        $a_2^X$ & $1.444380813060$ & $1.218553781780$\\
        $a_3^X$ & $0.479569465603$ & $0.661136493537$\\
        $a_4^X$ & $-2.286894741870$& $-1.405678925030$\\
        $a_5^X$ & $1.126632984980$& $-1.356418438880$\\
        $a_6^X$ & $1.250619843540$& $1.447029155340$\\
        $a_7^X$ & $-3.631020471590$& $4.235669735900$\\
        $a_8^X$ & $4.082217023790$& $-5.334045653410$\\
        $a_9^X$ & $0.504097346499$& $-2.916300520960$\\
        $a_{10}^X$ & $-5.085120460510$& $8.707403067570$\\
        $a_{11}^X$ & $3.967742543950$& $-5.706999943750$\\
        $a_{12}^X$ & $-0.981529071103$& $1.280814375890$\\\hline\hline
    \end{tabular}
    \caption{z-expansion proton form factor fitted parameters, taken from Ref.\cite{Ye:2017gyb}.}
    \label{tab:parameters}
\end{table}

\subsection{Setup 2: Lattice QCD Form Factors}
A second approach, which is also worth to consider, is a lattice QCD motivated computation of $a_\mu^{\mathrm{p-box}}$. In \cite{Alexandrou:2018sjm}, the Lattice data for the form factors can be parametrized using a simple dipole approximation:~\footnote{A z-expansion was performed as well, but no significant difference was found with respect to the simple dipole approximation.}
\eq{
G^{\{E,M\}}(Q^2)=G^{\{E,M\}}(0)/(1+Q^2/\Lambda)^2,\label{eq:formfactorslattice}
} where $\Lambda$ is related to the electric and magnetic radii by $\Lambda=12/\langle r_{\{E,M\}}^2\rangle$ and the normalization is $G_E(0)=1$ and $G_M(0)=\mu_p$. It is important to remark that this parametrization automatically fulfills the QCD-ruled asymptotic behavior for large values of $Q^2$. The numerical values required in eq.(\ref{eq:formfactorslattice}) are shown in table \ref{tab:parameters2}, where the normalization at null photon virtuality is automatically fulfilled for the electric form factor by setting $G_E^p(0) \to 1$. As explained in \cite{Alexandrou:2018sjm}, there is an underestimation of the electric r.m.s. radius, due to the slower decay of the electric form factor. Besides, the magnetic moment of the proton is undervalued, which could be caused by a combination of residual volume effects and multi-hadron contributions.

\begin{figure}[ht]
    \centering
    \begin{tabular}{cc}
       \includegraphics[scale=.34]{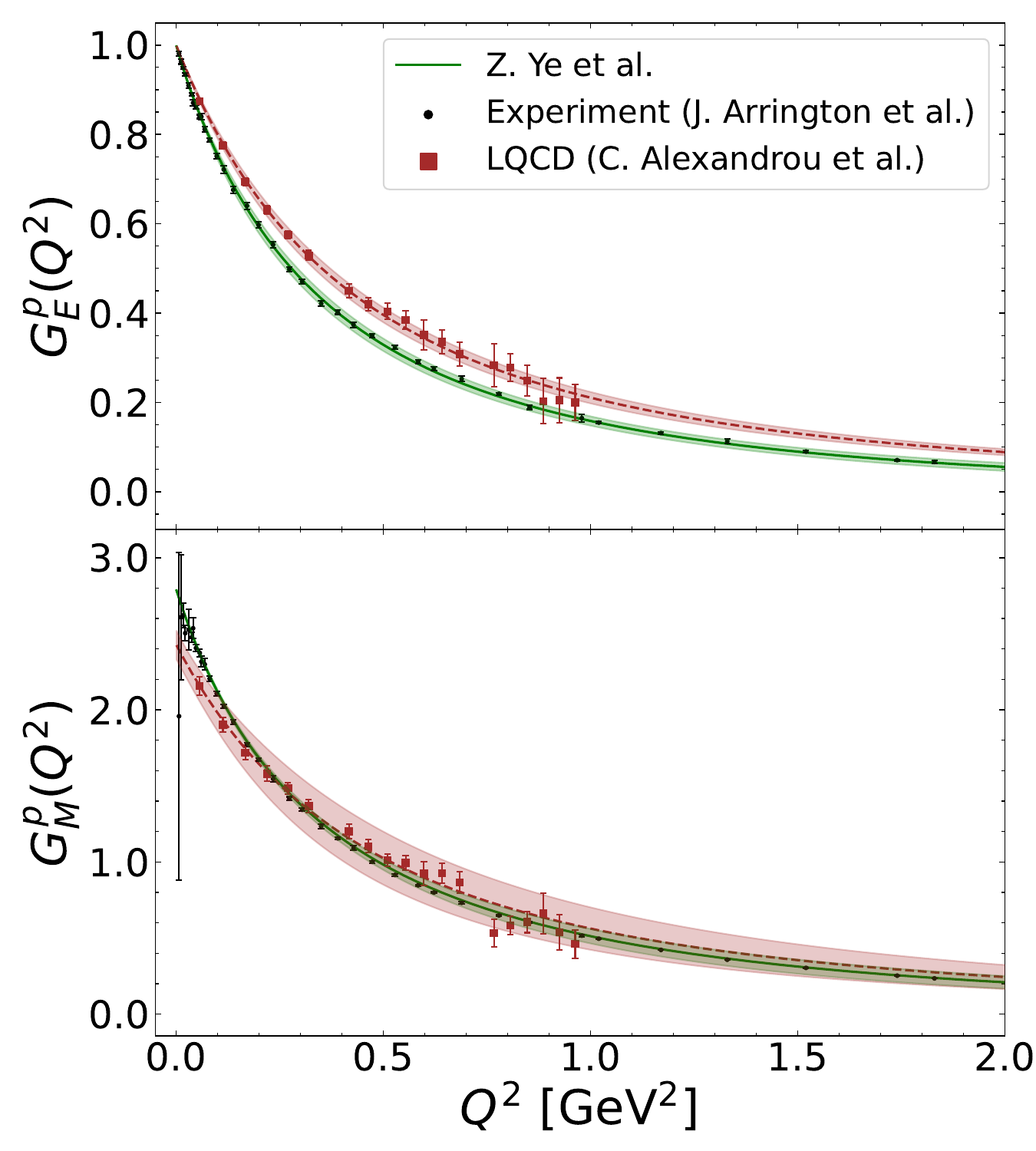} &
       \includegraphics[scale=.34]{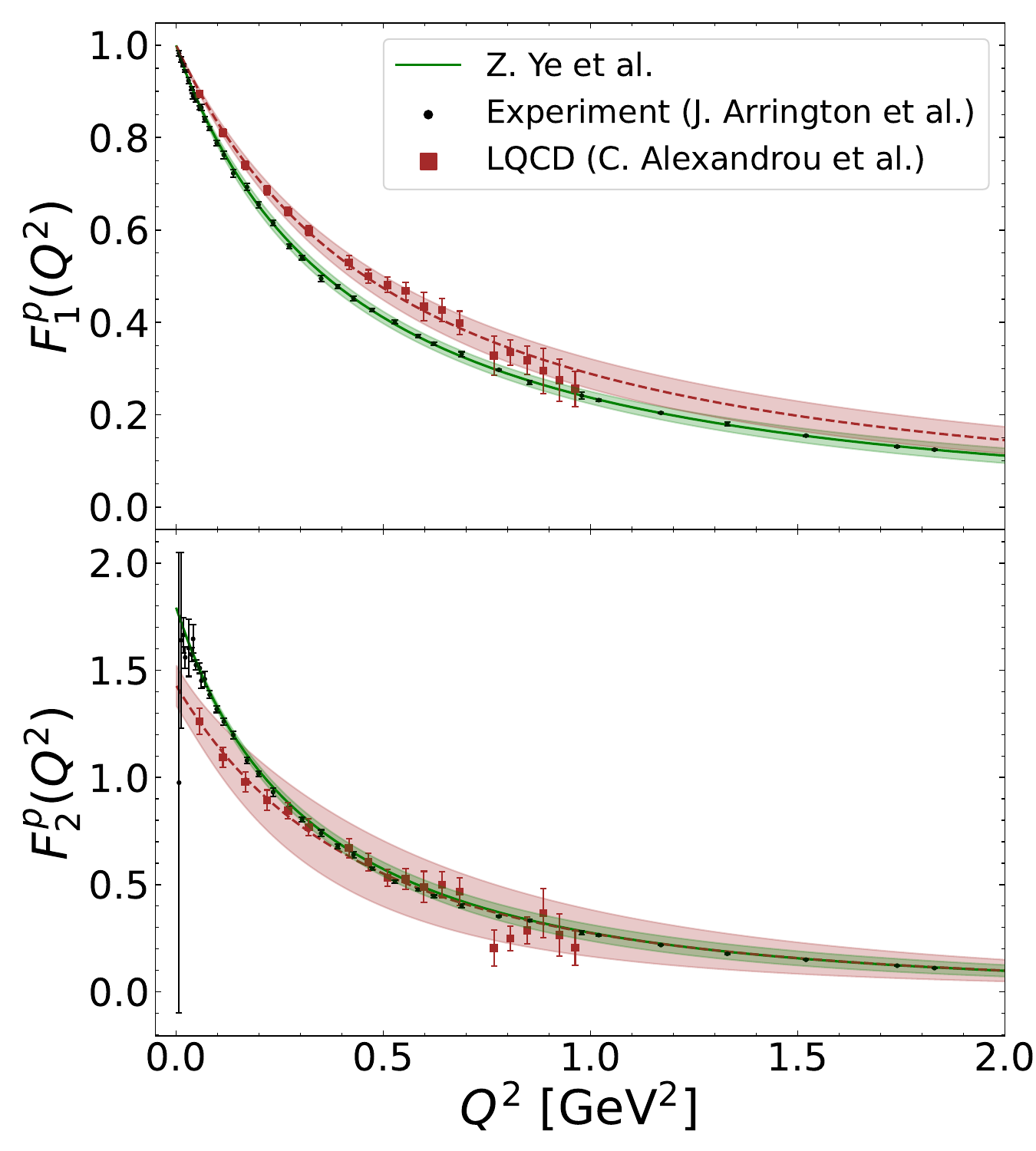} \\
        (a) & (b)
    \end{tabular}
    \caption{$G_E$ and $G_M$ ($F_1$ and $F_2$) proton form factors. In black we show the experimental points taken from Ref.\cite{Ye:2017gyb}, meanwhile, red dots correspond to the Lattice QCD results of Ref.\cite{Alexandrou:2018sjm}.}
    \label{fig:formfactors}
\end{figure}

\begin{table}[htp]
    \centering
    \begin{tabular}{c|c|c}\hline\hline
        \begin{tabular}{c}
        $\sqrt{\langle r^2_E\rangle}$[fm] 
        \end{tabular} & \begin{tabular}{c}
            $\sqrt{\langle r^2_M\rangle}$ [fm]
        \end{tabular} & \begin{tabular}{c}
            $\mu_p$
        \end{tabular}  \\\hline
        $0.742\pm0.013\pm0.023$ & $0.710\pm0.026\pm0.086$ & $2.43\pm0.09\pm0.04$\\\hline\hline
    \end{tabular}
    \caption{Numerical values of eq.(\ref{eq:formfactorslattice}) according to Ref.\cite{Alexandrou:2018sjm}. The uncertainties stand for the statistic and systematic errors, respectively.}
    \label{tab:parameters2}
\end{table}
As we show in Fig.\ref{fig:formfactors}, the z-expansion reported in \cite{Ye:2017gyb} is in good agreement with the data set of $G_M(Q^2)$ and $G_E(Q^2)$ from \cite{Arrington:2007ux} extracted from the world's data on elastic electron-proton scattering and calculations of two-photon exchange effects. We also compared the former fit and the one obtained from Lattice QCD results with a $N_f=2+1+1$ ensemble, reported in Ref.\cite{Alexandrou:2018sjm}. Finally, due to the small deviations between the two data sets, in the $Q^2<1\,\rm{GeV}^2$ region, it seems interesting to analyze both frameworks separately during the numerical evaluation of eq.(\ref{eq:master-formula}).  

\section{Proton-Box Contribution}
\label{sec:4}
In order to obtain the explicit $a_{\mu}^{\rm{HLbL}}$ contribution via the master integral, we implemented a numerical evaluation, using the VEGAS algorithm \cite{Lepage:2020tgj,Lepage:1977sw}. 

For a first consistency test of the integration method, we reproduced all previously well-known results, being in complete agreement with all of them ($\pi$-pole, $\pi$-box, c-loop, etc.). Specifically we corroborate that the use of the quark-loop scalar functions, without any form factors included, leads to the same result as the HME approximation for the proton case, getting a central value of $9.4\times10^{-11}$ compared to the HME estimation of $9.7\times10^{-11}$. 

Once the corresponding vector form factor $F_{1}(Q^2)$ is included in the analysis, we get the following results for the different setups described above:~\footnote{Using a different parametrization \cite{Alberico:2008sz} of the same data, as the setup 1, a result of $a_\mu^{\mathrm{p-box}}$=1.79(5)$\times 10^{-12}$ was found, consistent with the results using~\cite{Ye:2017gyb}.}
\eq{
a_{\mu}^{\mathrm{p-box}}=1.82(7)\times10^{-12}\quad \textbf{(Setup 1)},
}
\eq{
a_{\mu}^{\mathrm{p-box}}=2.38(16)\times10^{-12}\quad \textbf{(Setup 2)},
}
where both the systematic and statistic uncertainties were considered in order to estimate the error for each setup.

As previously discussed, the numerical suppression observed in the final result, relative to the HME approximation, can be directly attributed to the behavior of the kernel and form factors, as shown in Fig.~\ref{fig:kernels}. 
In this regard, we highlight the behavior as function of $\tau$ in three deferent regions:
\begin{itemize}
    \item Close to $\pm 1$, the $\sqrt{1-\tau^2}$ factor suppresses the values of the integral kernel.
    \item As $\tau$ increases, $Q_3$ does as well, and the $T_i$ decrease\cite{Colangelo:2017fiz}, causing the kernel to start diluting after its maximum value is reached, and to be almost negligible for positive values of $\tau$.
    \item The maximum values of the integration kernel appear in $\tau\in[-0.85,-0.65]$, as shown in the supplemental material. For this value of $\tau$ the relevant region of the kernel and the effect of the form factors in the numerical evaluation of eq. (\ref{eq:master-formula}) can be analyzed.
\end{itemize}
Indeed, the region where the integral kernel reaches its maximum contribution lies between the values of 0.1 and 0.01 of the form factor term, $F_1(Q_1^2)F_1(Q_2^2)F_1(Q_3^2)$ for all values of $\tau$.~\footnote{Despite Fig.~\ref{fig:kernels} results are presented for just three $\tau$ values. An .mp4 file showing the same behavior for all the $\tau$-range is added as supplemental material for this work.} Since in the HME, the proton is considered a point-like particle, the integrand of eq. (\ref{eq:master-formula}) is expected to be between 1 and 2 orders of magnitude smaller with respect to the structure-less case. 
This discrepancy between the peak locations of the kernels and form factors provides a clear and consistent explanation for the numerical integration results.

In the case of setup 1, both errors were computed for $F_1^p(Q^2)$ for each value of $Q^2$ as discussed in \cite{Ye:2017gyb}, and these $\Delta F_1^p(Q^2)$ were used for the error propagation of eq. (\ref{eq:master-formula}) considering the structure of eq. (\ref{eq:Scalar-functions}) in terms of the form factors. For the setup 2, a numerical computation of the Jacobian matrix of eq. (\ref{eq:master-formula}) within this setup was performed, and it was combined to obtain both the statistical and the systematic error by assuming a maximal correlation of the magnetic form factor parameters.\footnote{The uncertainty associated with the numerical integration method is subleading, of order $\mathcal{O}(10^{-15})$.}

Even though there is an underestimation of $\mu_p$ using a lattice QCD form factor, the slower decay of $G_E^p$ compared with the data-driven one, compensates for this, and it results in a higher $F_1^p(Q^2)$ for the lattice QCD result, as explained in Ref.\cite{Alexandrou:2018sjm} and visible in Fig.~\ref{fig:formfactors}. Consequently, the setup 2 result for $a_\mu^\mathrm{p-box}$ is larger than the one of the first setup. Therefore, in this work, we will adopt the data-driven $a_\mu^\mathrm{p-box}$ approximation as our central value, awaiting more precise lattice results anticipated in the near future.

\section{Conclusions}
\label{sec:5}
The hadronic light-by-light scattering is expected to soon dominate the theory uncertainty in $a_\mu$. Consequently, a detailed analysis of its various contributions has become an important task.

In this work we have computed a first approximation of the proton-box contribution to the HLbL piece of $a_\mu$. We discussed the corresponding proton form factor results, including a couple of different approaches, getting mutually consistent results from all of them. After implementing the master formula with the appropriate scalar functions, our data-driven analysis yields an estimated contribution of $a_{\mu}^{\mathrm{p-box}}=1.82(7)\times10^{-12}$. 

Finally, as already explained, a more precise result would require a full description of the scalar functions taking also into account the tensor vertex contribution. In such a way, the addition of the $F_{2}(Q^2)$ form factor -which is expected to be a subdominant contribution with respect to the vector term weighted by $F_{1}(Q^2)$- will improve the analysis, allowing us to also extend its application to any other baryon with well-characterized form factors, such as the neutron, for which the $F_1$ contribution trivially vanishes.

\section*{Acknowledgements}
Authors are really grateful with Dr.~Martin Hoferichter for enlightening correspondence and insightful comments. The authors also thank Prof. Giannis Koutsou for useful correspondence and Prof. Jhovanny Mejía \& Prof. Alberto Sánchez-Hernández for sharing their computational resources with us.
E.~J.~E., J.~M.~M.~ and D.~P.~S. are indebted to CONAHCYT funding their Ph.D. P.~R.~ is partly funded by CONAHCYT (México), with the support of project CBF2023-2024-3226 being gratefully acknowledged, and by MCIN/AEI/10.13039/501100011033 (Spain), grants PID2020-114473GB-I00 and PID2023-146220NB-I00, and by Generalitat Valenciana (Spain), grant PROMETEO/2021/071.

\appendix

\section{Fermion-Box Scalar Functions}
\label{App:Scalar-functions}

In this section, for completeness, we write the analytical expressions for the functions $I_i$ required for our analysis, cf. eq.~(\ref{eq:master-formula}), where the $\overline{\Pi}_i$ scalar functions enter. These had been obtained in Ref.\cite{Colangelo:2019uex} for a quark box-loop in terms of two Feynman parameters, $0\leq x\leq 1$ and $0\leq y\leq1-x$. We confirm the results in \cite{Colangelo:2017fiz}~\footnote{In ref.~\cite{Colangelo:2019uex} the $I_i$ functions multiply the $\hat{\Pi}_i$ functions. The relation between both bases is given in eq.~(2.22) of ref.~\cite{Colangelo:2017fiz}. Specifically, our $I_{1,3,5,9,10,12}$ correspond, respectively, to the $I_{1,4,7,17,39,54}$ in the tilded basis.}
\eq{
I_1=&-\frac{16x(1-x-y)}{\Delta^2_{132}}-\frac{16xy(1-2x)(1-2y)}{\Delta_{132}\Delta_{32}},\\
I_3=&\frac{32xy(1-2x)(x+y)(1-x-y)^2(q_1^2-q_2^2+q_3^2)}{\Delta^{3}_{312}}\nonumber\\
&-\frac{32(1-x)x(x+y)(1-x-y)}{\Delta_{312}^2}-\frac{32xy(1-2x)(1-2y)}{\Delta_{312}\Delta_{12}},\\
I_5=&-\frac{64xy^2(1-x-y)(1-2x)(1-y)}{\Delta_{132}^3},\\
I_9=&-\frac{32x^2y^2(1-2x)(1-2y)}{\Delta^2_{312}\Delta_{12}},\\
I_{10}=&\frac{64xy(1-x-y)((2x-1)y^2+xy(2x-3)+x(1-x)+y)}{\Delta^3_{132}},\\
I_{12}=&-\frac{16xy(1-x-y)(1-2x)(1-2y)(x-y)}{\Delta_{312}\Delta_{12}}\left(\frac{1}{\Delta_{312}}+\frac{1}{\Delta_{12}}\right),\\
}
where $\Delta_{ijk}=m^2-x y q_i^2-x(1-x-y)q_j^2-y(1-x-y)q_k^2$ and $\Delta_{ij}=m^2-x(1-x)q_i^2-y(1-y)q_j^2$. The rest of scalar functions, entering the master formula, can be obtained from $q_i$ permutations, as follows: 
\eq{\bar{\Pi}_2=&\,\mathcal{C}_{23}[\bar{\Pi}_1],\quad\bar{\Pi}_4=\mathcal{C}_{23}[\bar{\Pi}_3],\quad\bar{\Pi}_6=\mathcal{C}_{12}[\mathcal{C}_{13}[\bar{\Pi}_5]],\nonumber\\
\bar{\Pi}_7=&\,\mathcal{C}_{23}[\bar{\Pi}_5],\quad\bar{\Pi}_8=\mathcal{C}_{13}[\bar{\Pi}_{9}],\quad\bar{\Pi}_{11}=-\mathcal{C}_{23}[\bar{\Pi}_{12}],}
where the crossing operators $C_{ij}$ exchange momenta and Lorentz indices of the photons $i$ and $j$.




\bibliographystyle{JHEP.bst}

\bibliography{biblio.bib}
\end{document}